\documentclass{elsart}

\usepackage{graphicx}

\usepackage{amssymb}

\begin{document}

\begin{frontmatter}



\title{The diffuse neutrino flux from the inner Galaxy: \\  constraints from very high energy gamma-ray observations} 


\author[DIAS,KM3NET]{Stefano Gabici\corauthref{me}},
\author[MPIK,KM3NET]{Andrew M. Taylor},
\author[LEEDS,KM3NET]{Richard J. White},
\author[MPIK]{Sabrina Casanova}, 
\author[DIAS,MPIK,KM3NET]{Felix A. Aharonian};
\long\def\symbolfootnote[#1]#2{\begingroup\def\thefootnote{\fnsymbol{footnote}}\footnote[#1]{#2}\endgroup}
\symbolfootnote[2]{in cooperation with the KM3NeT Consortium}

\address[DIAS]{Dublin Institute for Advanced Studies,
31 Fitzwilliam Place, Dublin 2, IRELAND}
\address[MPIK]{Max-Planck-Institut f\"ur Kernphysik, 
             Postfach 103980, D-69029 Heidelberg, GERMANY}
\address[LEEDS]{School of Physics and Astronomy, University of Leeds, 
             Leeds LS2 9JT, UK}
\address[KM3NET]{Member of the KM3NeT Consortium. {\rm www.km3net.org}}

\corauth[me]{sgabici@cp.dias.ie}

\begin{abstract}
Recently, the MILAGRO collaboration reported on the detection of a diffuse multi-TeV emission from a region of the Galactic disk close to the inner Galaxy.
The emission is in excess of what is predicted by conventional models for cosmic ray propagation, which are tuned to reproduce the spectrum of cosmic rays observed locally.
By assuming that the excess detected by MILAGRO is of hadronic origin and that it is representative for the whole inner Galactic region, we estimate the expected diffuse flux of neutrinos from a region of the Galactic disk with coordinates $-40^{\circ} < l < 40^{\circ}$. 
Our estimate has to be considered as the maximal expected neutrino flux compatible with all the available gamma ray data, since any leptonic contribution to the observed gamma-ray emission would lower the neutrino flux.
The diffuse flux of neutrinos, if close to the maximum allowed level, may be detected by a km$^3$--scale detector located in the northern hemisphere. 
A detection would unambiguously reveal the hadronic origin of the diffuse gamma-ray emission.
\end{abstract}

\end{frontmatter}

\section{Introduction}
\label{intro}

Recently, the MILAGRO collaboration reported on the detection of multi-TeV diffuse gamma-ray emission from two regions in the Galactic plane: the Cygnus region (located at galactic longitude $65^{\circ} < l < 85^{\circ}$) \cite{MILAGROcygnus} and the portion of the inner Galaxy visible from the location of MILAGRO ($30^{\circ} < l < 65^{\circ}$) \cite{MILAGROinner}.
When placed alongside the earlier  
lower energy measurements by EGRET \cite{egretdiffuse}, such detections provide an interesting insight  
into both the spatial distributions of cosmic rays (CR) 
in the Galaxy and their energy spectrum. 

The diffuse GeV emission detected by EGRET along the galactic plane is believed to be the result of hadronic interactions between CRs and interstellar matter.
A strong point in favor of this hypothesis is the fact that the diffuse GeV emission correlates with the spatial distribution of the interstellar gas that constitutes the target for CR interactions \cite{egretdiffuse}.
The same argument might be extended also to the diffuse multi-TeV emission detected by MILAGRO, since both the Cygnus region and the inner Galaxy are characterised by an enhancement in the gas density.
In the case of the Cygnus region, the morphology of the multi-TeV emission correlates with the CO emission, which traces the gas density. This suggests again a hadronic origin also for the very high energy diffuse emission \cite{MILAGROcygnus}.
Similarly, for the region close to the inner Galaxy, the rather narrow latitude profile of the multi TeV-emission, concentrated around the dense Galactic disk, might suggest a hadronic interpretation of the gamma-ray data, though also a leptonic one seems feasible \cite{aa2000,galprophardelectrons,MILAGROinner}.

The expected hadronic gamma-ray emissivity from the Galactic plane can be calculated by solving the transport equation describing the propagation of CRs in the Galaxy and using the measured distributions of the gas density in the Galactic disk (see \cite{strongreview} for a review).
Conventional propagation models are tuned in order to match the CR spectrum observed locally \cite{galprop,dario,candia}.
These models reproduce quite fairly the observed level of the GeV emission (with the notable exception of the so called GeV-excess) \cite{galpropegret} but fail to reproduce the MILAGRO observations by about one order of magnitude in the case of the Cygnus region \cite{MILAGROcygnus} and a factor of $\sim 5$ for the inner Galaxy \cite{MILAGROinner}.
One possible solution of the problem is to assume that the average CR spectrum in the Galaxy is harder than the one we measure at Earth \cite{wolfendale,aa2000}. 
Another possibility is to assume a hard spectrum for the CR electrons \cite{aa2000,galprophardelectrons} or that the typical CR electron intensity in the Galaxy is significantly higher than that measured locally \cite{galpropegret}. This would make the inverse Compton contribution to the total gamma-ray emission dominate over the hadronic one at the multi-TeV energies relevant here.
Remarkably, both these scenarios (hadronic and leptonic) were proposed to solve the problem of the GeV--excess and can now be generalised to investigate their predictive power at multi-TeV energies.

Thus, the origin of the multi-TeV emission from the central Galactic plane region remains unclear.
Either leptonic or hadronic processes are believed to be the dominant contributors to the detected gamma-ray flux, 
with the possibility of a transition from a dominant hadronic contribution to a leptonic one in the GeV-TeV 
energy range \cite{galpropegret}.
In the case of the hadronic scenario, one would expect, as it is observed, a narrow extension in latitude of both the EGRET (see Fig.~3 from \cite{egretdiffuse}) and MILAGRO emission \cite{MILAGROinner}. In this case the extension of both GeV and multi-TeV emission is determined only by the gas distribution in the Galactic disk, which constitutes the target for proton--proton interactions.
In the leptonic scenario, the multi-TeV emission is produced via inverse Compton scattering by $\approx 100$~TeV electrons which can propagate over a distance of $\approx 100$~pc before being cooled by synchrotron and inverse Compton losses \cite{aa2000}. Thus, also in this case a narrow latitude distribution of gamma rays is expected, if the sources of cosmic ray electrons are concentrated around the galactic plane.
In this case, however, the extension of the gamma-ray emission would be energy dependent, since higher energy electrons are cooled faster and can propagate shorter distances \cite{aa2000}.  
Thus, an accurate comparison of the extension in latitude of the emission detected at GeV and multi-TeV energies is of crucial importance, and future observations with improved angular resolution by GLAST and HAWK might help in discriminating between the two scenarios. 

Since the production of gamma rays is accompanied by a corresponding neutrino flux only in the hadronic scenario, 
neutrino telescopes provide a unique opportunity to disentangle the origin of the multi-TeV radiation.
In this paper, we investigate the capability of km$^3$--scale neutrino detectors such as IceCube \cite{icecube} and KM3NeT \cite{km3net}, to detect the diffuse flux of neutrinos from the inner Galaxy and possibly reveal the origin of the diffuse emission detected by MILAGRO.


In the past, several estimates of the diffuse neutrino flux from the Galactic plane region have been 
published. The simplest of these approaches assumes that the CR distribution is constant across 
the Galactic disk \cite{stecker,berezinsky,ingelman}, whilst more sophisticated calculations assume a 
distribution of CR sources in the disk and obtain a steady state CR distribution by 
solving the diffusion equation \cite{dario,candia}. All of these approaches, however, result in relatively 
small neutrino fluxes and are below that required to be detectable by a km$^{3}$-size 
neutrino detector such as IceCube or KM3NeT, within one year of data acquisition. In these calculations, 
a CR spectral index of $\sim 2.7$ is generally assumed, or derived by solving the CR transport equation, since this matches the observed CR
spectrum measured at Earth. Such a steep spectrum, normalized to reproduce the diffuse emission
observed at GeV energies by EGRET, naturally results in a conservative estimate of the diffuse
neutrino flux at multi-TeV energies where km$^{3}$ neutrino telescopes are optimally sensitive. 
However, no a priori reason exists that such a CR spectrum is applicable throughout the Galaxy.
Indications exist for a departure from the observed local CR spectral index of $\sim E^{-2.7}$ 
in the diffuse gamma-ray emission observed by H.E.S.S. from the Galactic centre ridge \cite{H.E.S.S.ridge} and the GeV excess in the diffuse flux from the Galaxy observed in EGRET data, which might be due to a CR spectrum harder than the one observed at Earth \cite{wolfendale,aa2000}. 

In this work we adopt a phenomenological approach. 
We assume that the production of the multi-TeV diffuse flux observed by MILAGRO is entirely due to hadronic 
interactions. This model requires the assumption of a CR spectrum in the inner part of the Galaxy
that is different (harder) than the canonical 2.7 spectral index of the CRs observed at Earth.
Consequently we obtain a higher value for the multi-TeV neutrino flux from the same Galactic plane region
compared to previous canonical estimates. Any leptonic contribution to the multi-TeV diffuse emission
will lower the actual diffuse neutrino flux. Our calculations therefore provide an estimate of the maximal
neutrino flux that can be expected, under the constraint of remaining compatible with all the presently 
available gamma-ray data. 
Since the MILAGRO detection of the diffuse multi-TeV gamma-rays from the
inner Galaxy only covers the angular range $30^{\circ}<l<65^{\circ}$
\cite{MILAGROinner}, to apply this MILAGRO data point to the whole inner Galaxy with coordinates $-40^{\circ} < l < 40^{\circ}$ it
was assumed that the TeV flux detected by MILAGRO scales in a similar way to the
GeV flux detected by EGRET from the two regions. 

It is worth noticing that MILAGRO is operating at multi-TeV energies which is also the energy range where km$^3$-scale neutrino telescopes reach their best performances (see \cite{icecube_perf,km3net_perf} and Sec.~\ref{sec:telescopes}).
This implies that the MILAGRO observations are the most relevant gamma-ray observations to constrain the expected diffuse neutrino flux, since gamma rays and neutrinos are produced roughly at the same energy during CR hadronic interactions with the interstellar medium.

In Sec.~\ref{sec:gamma} we review the relevant gamma-ray observations that can constrain the expected neutrino flux from the inner Galaxy. We suggest different possibilities to fit the gamma-ray data in scenarios where the emission is dominated by hadronic CR interactions. 
In Sec.~\ref{sec:neutrinos} we use such fits to evaluate the expected flux of neutrinos from the inner Galaxy. This has to be considered as the maximal neutrino flux which is compatible to all the available gamma-ray measurements. Any leptonic contribution to the observed gamma-ray emission would lower the expectations. 
Our conclusions are provided in Sec.~\ref{sec:concl}.
The maximum level of the diffuse neutrino flux allowed by constraints from gamma-ray observations is within the detection capabilities of a km$^3$--scale neutrino telescope located in the northern hemisphere.
A detection would unambiguously reveal the hadronic origin of the multi-TeV emission detected by MILAGRO.


\section{Diffuse gamma rays from the inner Galaxy}
\label{sec:gamma}

The MILAGRO detection of a diffuse emission from both the Cygnus region ($65^{\circ}<l<85^{\circ}$) and the more central region of the Galaxy ($30^{\circ}<l<65^{\circ}$), represents more than seven years worth of data collection.
The large field of view of the detector ($\sim$2 sr) and its almost continuous data aquisition rate make the
instrument highly suitable for discovering large-scale diffuse gamma-ray signals which Air-Cherenkov telescope 
instruments would find more difficult to detect. The most significant of the MILAGRO diffuse flux detections
originates from the Cygnus region and is 8.6 $\sigma$ above the background \cite{MILAGROcygnus}. The central Galactic region was also 
detected with a smaller but still very robust significance of 5.1 $\sigma$ \cite{MILAGROinner}.
The median energy of the detected photons is in both cases 15 TeV.

\begin{figure}
\begin{center}
\includegraphics*[scale=.50, angle=-90]{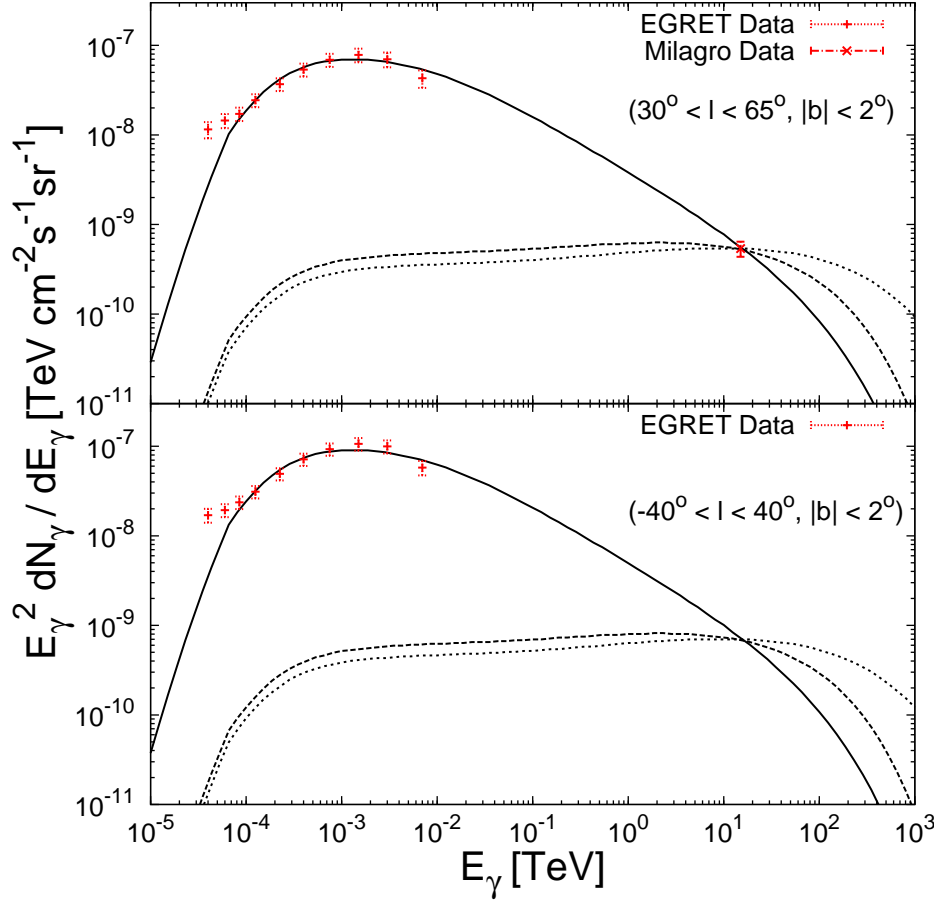}
\end{center}
\caption{\textbf{Top panel:} EGRET and MILAGRO data points for the region of the Galactic plane with coordinates $30^{\circ}<l<65^{\circ}$. The solid line represents a fit to both the sets of data due to $\pi^0$-decay gamma-ray emission from the CR spectrum described by Eq.~\ref{eq:broken}. The dotted and dashed lines represents the gamma-ray emission from a hard ($\propto E^{-2}$) CR spectrum that fits the MILAGRO point only. \textbf{Bottom panel:} EGRET data points for the inner Galaxy ($-40^{\circ}<l<40^{\circ}$). Lines are the same as in the top panel rescaled upwards of a factor $\sim 1.3$.
}
\label{fig:gamma}
\end{figure}
 
As here we are interested in the emission from the inner Galaxy, we plot in Fig.~\ref{fig:gamma} the available gamma-ray measurement for two distinct regions of the Galactic plane. In the upper panel of Fig.~\ref{fig:gamma} we show the MILAGRO point relative for the region with Galactic coordinates $30^{\circ} < l < 65^{\circ}$ and $-2^{\circ} < b < 2^{\circ}$, together with the EGRET points for the same region. 
This is the region within the field of view of MILAGRO which is the closest to the Galactic centre.
As mentioned in Sec.~\ref{intro} and demonstrated in \cite{MILAGROinner}, conventional models assuming that the gamma-ray emission is dominated by hadronic CR interactions, and that the CR spectrum has the canonical slope $\sim 2.7$ everywhere in the Galaxy fail to fit simultaneously the EGRET and MILAGRO data by roughly a factor of 5.
A good fit to all the gamma-ray data can be obtained by assuming that the CR spectrum in the region considered here is not a single power law but rather a broken power law:
\begin{equation}
\label{eq:broken}
J(E)=J_{0} ~~ \frac{E^{-\alpha}}{\left[1+(\frac{E}{E_{*}})^{\delta}\right]} ~~ e^{(-E/E_{\rm cut})} ~ ,
\end{equation} 
with slope $\alpha = 2$ and $\alpha + \delta = 2.7$ below and above $E_* = 80$ GeV respectively.
An exponential cutoff is added at an energy of $E_{{\rm cut}} = 1$ PeV, close to the energy of the CR knee.
$J_0$ is a free normalization parameter, determining the total energy in the form of CRs.
A physical justification of such a spectral shape has been provided in \cite{aa2000}, where the authors suggested to interpret $E_*$ as the energy where the escape of CRs from the Galaxy changes from convective to diffusive.
Thus, at low energies (below $E_*$), where convection in the galactic wind dominates, the shape of the CR spectrum mimic the injection spectrum ($\approx E^{-2}$), while at higher energies (above $E_*$) diffusive escape of particle from the Galaxy steepens the spectrum to its canonical value $\sim E^{-2.7}$.

The gamma-ray emission resulting from proton--proton interactions between CRs with the spectrum considered in Eq. \ref{eq:broken} and interstellar gas has been calculated using the parameterization given in \cite{kelner} and is plotted as a solid line in Fig. \ref{fig:gamma} (top panel)\footnote{The best fit parameters slightly change if a different parameterization for the production of gamma rays is adopted (e.g. \cite{dermer,aa2000,kamae}) since these parameterizations treat the production of low energy pions in proton--proton interactions in a different way. However, at the high energies which are most relevant for this paper all the approaches converge to the same spectrum, and the choice of the parameterization will not affect our final results.}.
The curve fits both the EGRET and MILAGRO data, with the exception of the data points at energies below $\approx 100$ MeV, where electron Bremsstrahlung is well known to be the dominant mechanism for gamma-ray production in the Galactic disk \cite{aa2000,galpropegret}.

A different possibility to explain the excess of multi-TeV diffuse emission above the conventional models without invoking leptonic contributions is to assume the existence of a hard hadronic component in the gamma-ray spectrum that adds up to the canonical $\sim E^{-2.7}$ component and dominates the total emission above $\approx$ TeV energies.
Such a hard gamma-ray spectra can be produced in different ways. 
One possibility is to have a number of unresolved CR sources within the emitting region, as suggested in \cite{MILAGROcygnus}. In this case, the CR spectrum inside the sources is not steepened yet by diffusion in the interstellar medium and might thus contribute significantly to the multi-TeV emission.
Hard gamma-ray spectra can also be produced by propagation effects when relatively young CR sources exist in an inhomogeneous interstellar medium which hosts molecular cloud complexes \cite{aa1996,me}.
This is a consequence of the energy-dependent nature of CR propagation away from their sources in the interstellar medium. Under certain conditions, an energy-dependent diffusion coefficient allows the formation of hard CR spectra around the sources, whose $\pi^0$-decay gamma-ray emission can be amplified by the presence of a massive molecular cloud.
Thus, a population of CR accelerators (such as supernova remnants) might be responsible for the whole gamma-ray emission detected by MILAGRO from both the Cygnus and the inner Galactic region, which host dense molecular cloud complexes.

In this paper we adopt a phenomenological approach, not focussing on any particular scenario for the production of the observed multi-TeV gamma rays. Being interested in obtaining the maximal flux of diffuse neutrinos  we consider for further calculations two different CR spectra, which are felt to cover the set of plausible scenarios for a hadronic origin of the diffuse gamma-ray emission detected by MILAGRO: \textit{i)} the broken power law described by Eq.~\ref{eq:broken}, which fits simultaneously the EGRET and MILAGRO data and \textit{ii)} a hard CR spectrum of the form:
\begin{equation}
\label{eq:hard}
J(E)=J_{0} ~ E^{-2} ~ e^{(-E/E_{\rm cut})} ~ ,
\end{equation}
which fits the MILAGRO point and is represented in Fig.~\ref{fig:gamma} (top panel) by the dashed and dotted lines, characterized by two different values for the position of the high energy cutoff of 1 and 5 PeV respectively. 
The hard CR spectrum described by Eq.~\ref{eq:hard} provides a negligible contribution to the low energy ($\approx$ GeV) gamma-ray emission detected by EGRET, that can be explained by conventional models that assume a steep CR spectrum \cite{egretdiffuse,galpropegret}.

In order to estimate the diffuse neutrino flux from the inner Galaxy we proceed as follows.
We call here inner Galaxy the region of galactic coordinates $-40^{\circ} < l < 40^{\circ}$ and $-2^{\circ} < b < 2^{\circ}$. This is the region for which the diffuse emission detected by EGRET shows the highest brightness (see Fig.~2a from \cite{egretdiffuse}). Thus, if the emission has, as it is believed, a hadronic origin, the expected neutrino flux is also expected to be enhanced within this region. The missing piece of information is the gamma-ray flux from the inner Galaxy at multi-TeV energies, which is the most relevant to constrain the neutrino emission.
This flux is not available since the MILAGRO observations cover only a fraction of the inner Galaxy.
We assume here that the enhancement detected by MILAGRO in the region of coordinates  $30^{\circ} < l < 65^{\circ}$, which partially overlaps with the region we consider, is representative for the whole inner Galaxy and that we can rescale it following the EGRET gamma-ray brightness profile.
In Fig.~\ref{fig:gamma} (bottom panel) we show the EGRET data points for the inner Galaxy, together with the three curves adopted in the upper panel to fit the gamma-ray data.
All the curves have been multiplied by 1.3 in order to follow the spatial variation of the EGRET flux.
This assumption is equivalent to saying that GeV and multi-TeV CRs 
are distributed in the inner Galaxy in the same way.
It has to be kept in mind that the factor 1.3 takes into account of both the spatial distribution of CRs and the increase of the gas density toward the centre of the Galaxy.

Before proceeding to estimate the expected neutrino flux, it is important to note that upcoming observations by GLAST will improve our knowledge of the Galactic diffuse gamma-ray emission and will extend the observed energy range up to $\sim 100$~GeV.
Similarly, next generation instruments like HAWK will survey the gamma-ray sky in the energy range between $\sim 1$ and 100~TeV.
The improved sensitivity and better coverage of the energy range will allow a more accurate fit to the data and consequently reduce the uncertainties in the estimate of the neutrino flux.

\section{Neutrino flux from the inner Galaxy}
\label{sec:neutrinos}

In this section, after briefly reviewing the capabilities of km$^3$--scale neutrino telescopes, we estimate the diffuse flux of neutrinos from the inner part of the Galaxy ($-40^{\circ}<l<40^{\circ}, -2^{\circ}<b<2^{\circ}$) by assuming that the CR spectra that we used in the previous section to fit the MILAGRO data are representative for the whole inner Galaxy.
These are the CR spectra that produce the $\pi^0$--decay emission plotted in the bottom panel of Fig.~\ref{fig:gamma}.

\subsection{Capabilities of neutrino telescopes}
\label{sec:telescopes}

\begin{figure}
\begin{center}
\includegraphics*[scale=.26, angle=-90]{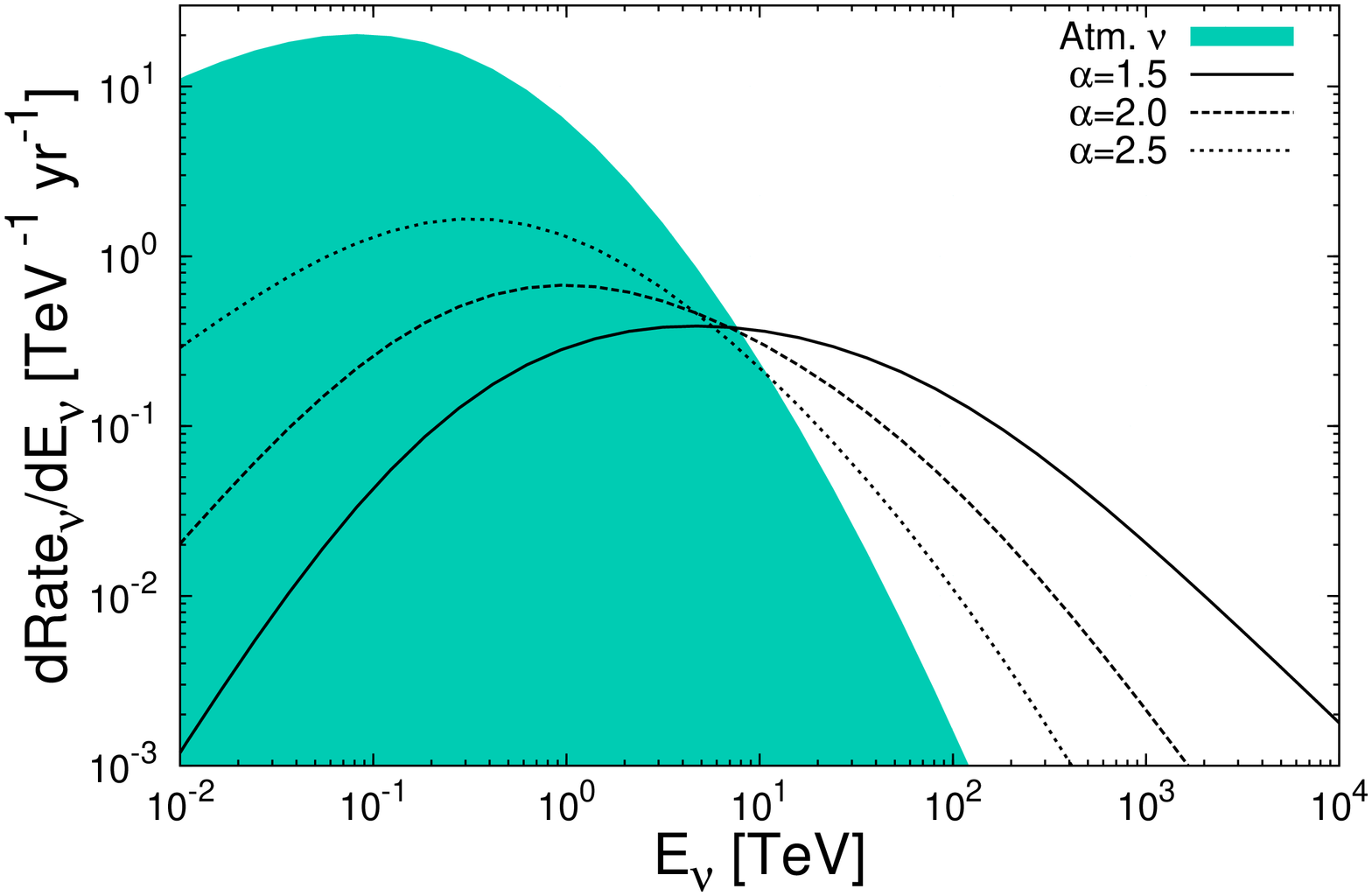}
\hspace{0.1 cm}
\includegraphics*[scale=.26, angle=-90]{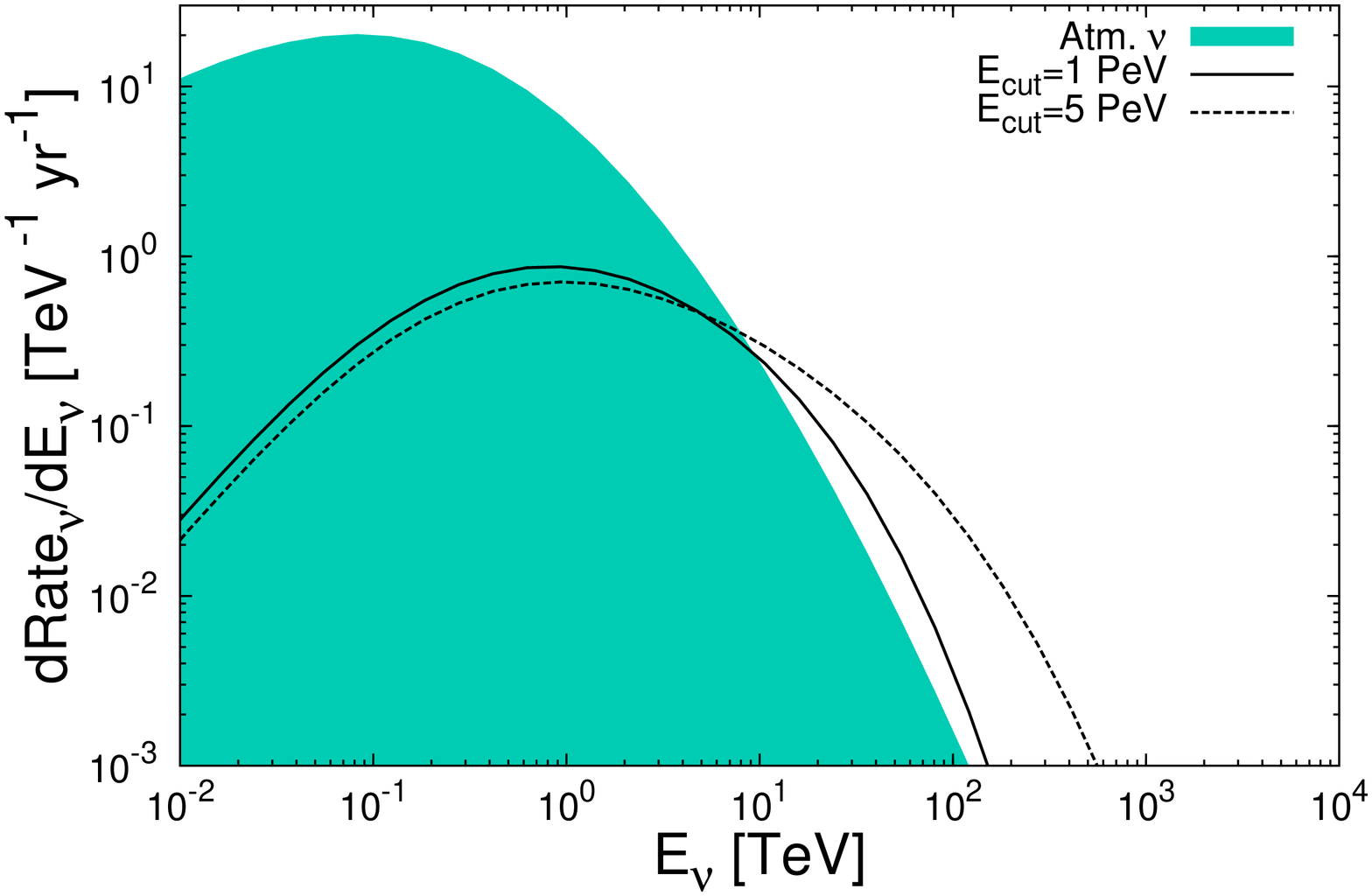}
\end{center}
\caption{Differential detection rates for neutrino sources with flux $F(>1~TeV) = 10^{-11} \nu$ cm$^{-2}$ s$^{-1}$. The dashed region represents atmospheric neutrinos.  \textbf{Left panel:} Solid, dashed and dotted lines represent the count rates for neutrino sources with power law spectrum with slope 1.5, 2 and 2.5 respectively and high energy cutoff at $E \gg 1{\rm PeV}$. \textbf{Right panel:} Count rates for neutrino sources with $\propto E^{-2}$ spectra and exponential cutoff at $E_{\rm cut} =$ 10 and 100 TeV (solid and dashed line respectively).
}
\label{fig:richard}
\end{figure}

As a rule of thumb, km$^3$--scale neutrino telescopes can detect a persistent and point--like source at a flux level of $\approx 10^{-11} \nu$ cm$^{-2}$ s$^{-1}$ after a few years of continuous observations.
This flux is the total flux integrated above $\approx 1$ TeV and roughly corresponds to the observed flux of the Crab nebula in gamma rays.
An accurate determination of the detection rate from a Crab-like source of neutrinos can be obtained by considering the telescope's effective area and the spectrum of the neutrinos received by the source. The rate obtained in this way has to be compared with the detection rate of atmospheric neutrinos, which constitute the dominant background.
All these aspects will determine which is the optimal energy range for the detection of astrophysical neutrinos.

As an illustrative example, we consider here a hypothetical point--like and steady source of neutrinos with differential flux: 
\begin{equation}
J(E)=J_{0}\left(\frac{E}{\rm TeV}\right)^{-\alpha}e^{(-E/E_{\rm cut})}
\end{equation}
with $J_{0}$ normalised such that the integrated flux above 1 TeV is 10$^{-11}$~cm$^{-2}$~s$^{-1}$.
$\alpha$ and $E_{\rm cut}$ are free parameters.
For the effective area we used the one provided in \cite{CDR,thesis,kappes} for the KM3NeT detector. Being located in the northern hemisphere, this telescope will be able to observe the inner part of the Galaxy and thus is the more relevant for this paper.
In the calculations, we used a convenient fit to the effective area which reads:
\begin{equation}
\label{eq:area}
A_{{\rm eff}}(E) = 10^{4.7} ~ E^{3.4} ~ \left( \frac{0.24}{0.24+E} \right) ~ \left( \frac{0.31}{0.31+E} \right) ~ \left( \frac{37}{37+E} \right) ~~ {\rm cm}^2
\end{equation}
where $E$ is the neutrino energy in TeV.
The product between the effective area and the neutrino flux within one angular resolution element  results in the 
expected differential detection rate by KM3NeT, and it is shown in Fig.~\ref{fig:richard}.
Rates are defined as the number of neutrinos detected after one year of exposure.

In the left panel of Fig.~\ref{fig:richard} we demonstrate the effect of changing the spectral slope of the neutrino source. The solid, dashed and dotted curves refer respectively to $\alpha =$ 1.5, 2 and 2.5, that cover the most plausible spectra for neutrino sources. The position of the exponential cutoff is $E_{{\rm cut}} \gg 1 {\rm PeV}$.
In the right panel we fix $\alpha =$2 and we change the energy at which the spectrum cuts off.
Solid and dashed lines refer to $E_{\rm cut} =$ 10 and 100 TeV respectively.
In both the plots the shaded region represents the level of the atmospheric neutrino background, assumed to follow the spectrum predicted in \cite{Volkova}. The resulting
atmospheric background rate is shown to be consistent with
\cite{Barr,Bartol,Honda}.
The adopted atmospheric background assumed a zenith angle of $\sim 70^{\circ}$. This is the effective zenith angle of the inner Galactic region we are interested in ($-40^{\circ}<l<40^{\circ}, -2^{\circ}<b<2^{\circ}$), obtained by integrating the atmospheric neutrino flux over one day from source regions below the horizon.
The energy resolution of the detector is assumed to be constant
across the entire energy range. It is accounted for by convolving the
contents of each energy bin with a Gaussian distributed in logarithmic
energy space. The Gaussian has an RMS in the difference between the
natural logarithm of the reconstructed and true neutrino energies of
0.3. 
The angular
resolution of the detector is given by a parameterisation of the RMS
difference between the reconstructed neutrino direction and the true
neutrino direction from the same simulations used to produce the
effective area curve \cite{CDR,thesis}, and above 1~TeV drops below 0.2$^\circ$. In
these generic plots it is assumed the source is point-like, and both
the signal and background neutrino fluxes are integrated across a cone
of optimal opening angle 1.58 times the angular resolution.

It is evident from Fig.~\ref{fig:richard} that the prospects for km$^{3}$ size neutrino telescopes to detect such fluxes are optimal in the 
10-1000 TeV energy range, where the signal flux is well above that due to atmospheric background events 
and the effective area grows sufficiently quickly with energy for detection to be possible. This result
comes with the caveat that the cutoff energy of the flux, $E_{\rm cut}$, sits at sufficiently high energies ($>$100 TeV). 
Thus the best targets for neutrino telescopes are sources exhibiting a hard spectrum which extends up to at least hundreds of TeV.
It is for this reason that the detections of multi-TeV gamma-ray discrete sources and diffuse emission by MILAGRO are of such relevance for future prospects for the detection of the first high energy astrophysical neutrinos from the Galaxy.

Qualitatively, similar conclusions can be reached also in the case of diffuse sources, the only difference being that we have to integrate both the signal and the atmospheric background over the extension of the sources.
In the next section we will focus on this issue, by considering the expected neutrino emission from the inner part of the Galactic disk.

\subsection{Neutrinos from the inner Galaxy: detectability for km$^3$--scale detectors}

\begin{figure}
\begin{center}
\includegraphics*[scale=.26, angle=-90]{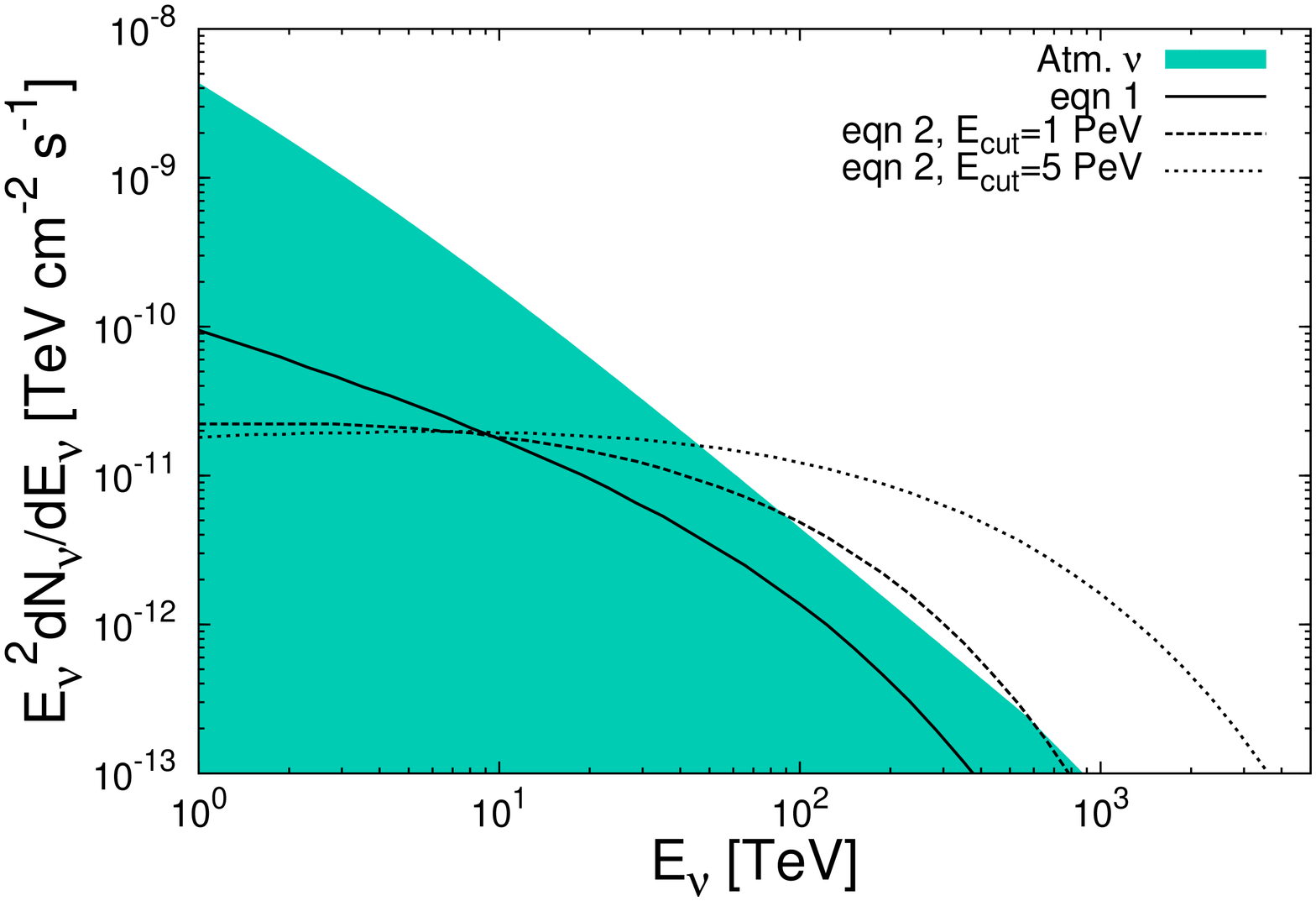}
\hspace{0.1cm}
\includegraphics*[scale=.26, angle=-90]{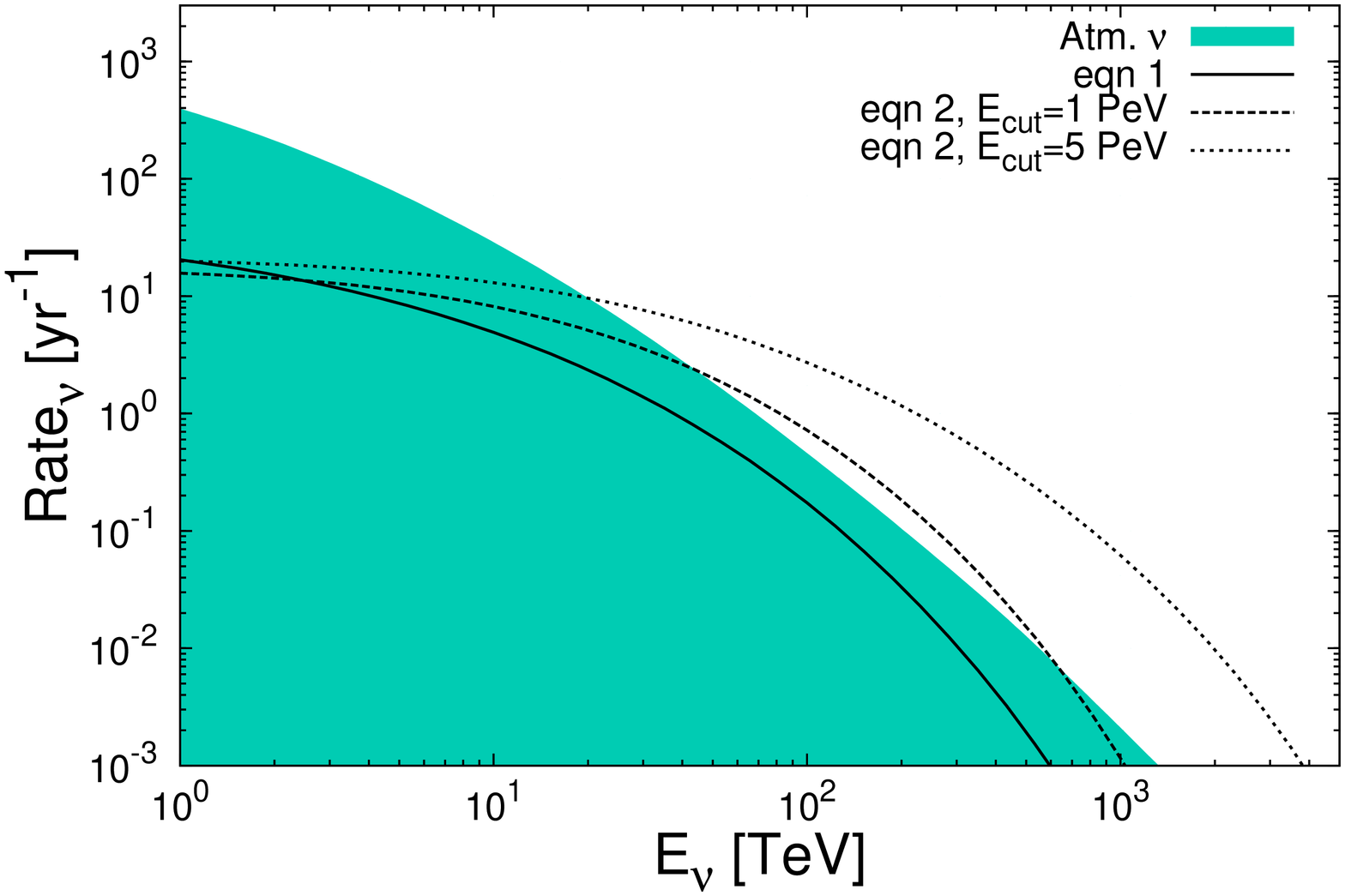}
\end{center}
\caption{\textbf{Left panel:} neutrino fluxes from the inner Galaxy. Solid, dashed and dotted lines refer to an underlying CR spectrum described by Eq.~\ref{eq:broken} and Eq.~\ref{eq:hard} with $E_{\rm cut} =$ 1 and 5 PeV respectively. The flux of atmospheric neutrinos is shown as a shaded region. \textbf{Right panel:} Detection rates integrated above an energy $E$ in a km$^3$ neutrino telescope. Curves have the same meaning as in the left panel.
}
\label{fig:nu}
\end{figure}



The left panel of Fig.~\ref{fig:nu} shows our predictions for the diffuse flux of neutrinos expected from the inner Galaxy, defined by the coordinates $-40^{\circ}<l<40^{\circ}, -2^{\circ}<b<2^{\circ}$.
Neutrino spectra are calculated following \cite{kelner} and are the sum of all the neutrino flavours produced in proton--proton interactions (two muon neutrinos plus one electron neutrino) divided by three to take into account neutrino oscillations (maximal mixing).
The adopted parent proton spectra are the same we used to fit the MILAGRO data (see Sec.~\ref{sec:gamma}) and then extrapolated to the whole inner Galaxy, namely, we used the same CR proton spectra that produce the gamma-ray emission plotted in the bottom panel of Fig.~\ref{fig:gamma}.

The solid line represents the neutrinos produced by CRs with spectrum described by the broken power law in Eq.~\ref{eq:broken}, while dashed and dotted lines refer to a hard CR spectrum like in Eq.~\ref{eq:hard}, with exponential cutoff at energies $E_{\rm cut}=$ 1 and 5 PeV respectively.
The shaded region represents the atmospheric neutrinos, integrated over the whole of the considered region, which has an extension of 0.097 steradians.

For the case of the broken power law the atmospheric neutrinos dominates over the signal for all neutrino energies. On the other hand, if a hard spectrum is considered, and the exponential cutoff in the CR spectrum is significantly above 1 PeV, the neutrinos from CR interactions in the inner Galaxy start to dominate over the atmospheric background above an energy of a few tens of TeV. 


In order to investigate the capability of a neutrino telescope to detect the diffuse emission we convolved the neutrino fluxes with the effective area (Eq.~\ref{eq:area}) to obtain the integral detection rate.
The rates above a given energy $E$ for one year of exposure are shown in the right panel of Fig.~\ref{fig:nu}.
We neglect here the effect of a finite energy resolution of the detector. 
Line types are the same as in the left panel of Fig.~\ref{fig:nu}.
It is evident that the detection of diffuse neutrinos from the inner Galaxy becomes most
feasible for neutrino energies above a few tens of TeV, where the signal becomes comparable or above the background counts rate. 
Above 10 TeV a rate of 5, 9 and 15
neutrinos is expected after a single year of exposure for the solid, dashed and dotted curve respectively, against a background of about 28 neutrinos
per year. 
For the most optimistic case (dotted line in Fig.~\ref{fig:nu}) the neutrino signal becomes equal to the background at an energy of 20 TeV with about 10 events per year.

The visibility of the inner Galaxy
must also be considered, which for a detector
located in the Mediteranean at 42$^\circ$50'$N$, 6$^\circ$10'$E$, is
visible for $\approx$ 70\% of the year.  
The rates shown in Fig.~\ref{fig:nu} have to be considered as the number of neutrinos after one year of exposure, which translates in $\approx 1.5$ yr of physical time.

\section{Discussion and conclusions}
\label{sec:concl}

In this paper we presented an estimate of the neutrino flux from the inner Galaxy, defined as the region of coordinates $-40^{\circ} < l < 40^{\circ}$ and $-2^{\circ} < b < 2^{\circ}$.
Our calculations rely on the following assumptions:
\begin{enumerate}
\item{the diffuse gamma-ray emission detected at multi-TeV energies by MILAGRO from the region of the Galactic plane of coordinates $30^{\circ}<l<65^{\circ}$ is representative for the whole inner Galaxy;}
\item{the diffuse galactic GeV and multi-TeV emission detected by EGRET and MILAGRO respectively follows the same spatial distribution;}
\item{the diffuse gamma-ray emission is entirely of hadronic origin.}
\end{enumerate} 
These assumptions allow to obtain an almost model independent estimate for the diffuse flux of neutrinos from the inner Galaxy. 
Such an estimate has to be considered as the maximal expected flux of neutrinos compatible with gamma ray observations, since any leptonic contribution to the observed gamma ray emission would lower our expectations.

The maximal expected neutrino flux is within the detection capabilities of a km$^3$-scale neutrino telescope located in the northern hemisphere. 
For the most favorable assumptions made in Sec.~\ref{sec:neutrinos}, we expect from the inner Galactic region 15 neutrinos above 10 TeV per year of exposure against a background of 28 atmospheric neutrinos. At 20 TeV the signal equates the background with about 10 events per year.
The detection of these neutrinos would constitute an unambiguous proof of the hadronic origin of the gamma-ray emission detected by MILAGRO.
Finally, it is important to stress that MILAGRO is operating in an energy range similar to the one where neutrino telescopes reach their best performances.
This makes MILAGRO and future instruments like HAWK, the ideal gamma-ray detectors for multi--messenger studies of sources of astrophysical neutrinos.

\section*{Acknowledgments}

SG acknowledges the support of the European Community under a Marie Curie Intra--European Fellowship.
AT acknowledges a research stipendium.
RW acknowledges support through the KM3NeT Design Study (supported by EU FP6 under contract no. 011937).



\end{document}